\documentclass[twocolumn,aps, prl]{revtex4}

\usepackage{graphics}
\usepackage{graphicx}
\usepackage{dcolumn}
\usepackage{bm}
\usepackage{color}

\newcommand{\av}[1]{\langle#1\rangle}

\newcommand{\de}{\partial}
\newcommand{\sech}[1]{\textrm{sech}\left(  #1\right)}
\newcommand{\eq}[2]{\begin{equation} \label{#1} #2 \end{equation}}

\newcommand{\CS}{CS$_{2}$ }
\newcommand{\schr}{Schr\"odinger }

\begin{document}


\title{Linearons: highly non-instantaneous solitons in liquid-core photonic crystal fibers}
\author{Claudio Conti$^1$, Markus Schmidt$^2$, Philip St.J. Russell$^2$ and Fabio Biancalana$^{2}$}
\affiliation{$^1$CNR-ISC, Department of Physics, University La Sapienza, P.le A.Moro 2, 00185 Rome, Italy\\$^2$ Max Planck Institute for the Science of Light, 91058 Erlangen, Germany}
\date{\today}

\begin{abstract}
The nonlinear propagation of light pulses in liquid-filled photonic crystal fibers is considered.
Due to the slow reorientational nonlinearity of some molecular liquids, the nonlinear modes propagating inside such structures can be approximated, for pulse durations much shorter than the molecular relaxation time, by {\em temporally} highly-nonlocal solitons, analytical solutions of a {\em linear} \schr equation. The physical relevance of these novel solitary structures, which may have a broad range of applications, is discussed and supported by detailed numerical simulations.
\end{abstract}

\maketitle

{\it Introduction ---} Highly nonlocal solitons have been originally introduced as an ``accessible'' toy-model for describing self-trapped optical beams \cite{Snyder97}; but this was followed by experimental demonstrations  \cite{ContiPRL04}, also driven by
early works on plasma physics, Bose-Einstein condensation and dissipative systems \cite{Litvak75,Pecseli80,Turitsyn85,Segev92,Vanin94,Akh98,Parola98,Perez00},
which unveiled the fundamental role of nonlocality in spatially self-trapped waves. Indeed, nonlocality allows stabilization with respect to collapse
and the existence of a rich class of propagation-invariant waves \cite{Bang02,Ouyang09}; in addition,
applications such as light-steering and all-optical logic gates \cite{Peccianti04nature} and
in soft-matter and thermal liquids \cite{Rotschild05,Conti05PRL,Kartashov07} have been demonstrated.
Recently, insights from spatially nonlocal nonlinear waves also emerged in the temporal domain \cite{shock2010}.
However, the relevance of temporal nonlocality is largely limited by the unavoidable instantaneous Kerr effect, as, e.g., for silica glass in fiber optics,
where ``nonlocal'' Raman-like terms, although leading to important consequences as the Raman self-frequency shift (RSFS) of solitons \cite{agrawalbook}, can be considered as a small perturbation.
Recent fabrication advances, however, open up innovative perspectives. Indeed, micro-structured photonic-crystal fibers (PCFs) may be fabricated with a central hole
filled by a material displaying non-instantaneous response as, e.g., molecular liquids with large reorientational effects
\cite{herrmannjosab}, with instantaneous nonlinearities acting as small perturbations: exactly the opposite situation of the silica glass.

In this Letter we show that a completely novel class of solitary wave exists in these new fibers, which are described by an essentially {\em linear} model.
These {\em linearons} are shown to display truly remarkable properties,
such as robustness with respect to noise perturbations and collisions, and emission of phase-matched dispersive waves - phenomena that are typically associated to purely nonlinear temporal waves only.
In addition they are shown to be not affected by the RSFS if the response time of the molecular reorientation is slow enough.
As a result, these highly noninstantaneous fibers can be used in a new variety of optical devices and applications, also including
quantum information processing.

{\it Liquids with reorientational nonlinearities ---}
Suitable liquids with large reorientational nonlinearities and comparably small Kerr contributions are those composed of small molecules. Good candidates are solvents typically used for spectroscopy. These liquids, which are available in extremely high purity, show large transparency windows, ranging from 500 nm to 3 $\mu$m \cite{cook}, high optical damage threshold and comparably long reorientational response times.
Experimental implementation requires the liquid to possess certain physical properties in addition to the nonlinear optical ones. An in-depth survey shows that best potential candidates are nitrobenzene (C$_{6}$H$_{5}$NO$_{2}$), toluene (C$_{7}$H$_{8}$) and carbon disulfide (CS$_{2}$), which all possess small cigar-shaped molecules with comparable low Kerr nonlinearity (C$_{6}$H$_{5}$NO$_{2}$: 671$\times$10$^{-16}$cm$^{2}$/W, C$_{7}$H$_{8}$: 168$\times$10$^{-16}$cm$^{2}$/W, CS$_{2}$: 340$\times$10$^{-16}$cm$^{2}$/W; all at 1064 nm \cite{sutherland}) and refractive indices larger than silica. From a practical point of view, both nitrobenzene and toluene have relatively low vapor pressure at room temperature and hence are easy to handle. CS$_{2}$, in contrast, is volatile, but has the best transparency properties due to its simpler molecular composition. To illustrate our findings we have performed calculations for CS$_{2}$, which is a common reference material in spectroscopy and thus all of its optical and physical properties are extremely well documented \cite{cs2}. However, we have to stress that our results are of general character and are not restricted to one specific liquid.

{\it Linearon equation ---}  Our starting point is the GNLSE written in dimensionless form:
\eq{eq1}{i\de_{z}A+\frac{1}{2}s\de_{t}^{2}A+A\int_{-\infty}^{+\infty}R(t-t')|A(t')|^{2}dt'=0,} where $R(t)=(1/T)\exp(-t/T)\Theta(t)$ is the response function of the \CS reorientational nonlinearity, $T=1.68$ psec is the medium response time \cite{cs2}, $\Theta(t)$ is the Heaviside function that ensures causality, $A(z,t)$ is the electric field envelope, $s=+1$ and $s=-1$ denote anomalous and normal group velocity dispersion (GVD) respectively, $z$ is the dimensionless propagation coordinate along the fiber, and $t$ is the dimensionless time coordinate. The response function $R(t)$ must be normalized to one: $\int R(t)dt=1$. We further define the moments of the response function: $\av{T^{(m)}}\equiv\int t^{m}R(t)dt$.

In the well-known limit (valid for materials such as silica, for instance, see \cite{agrawalbook}) of first moment much shorter than the pulse duration $t_{0}$, i.e. $T\equiv T^{(1)}\ll t_{0}$, one can use the reversibility property of the convolution integral
$\int_{-\infty}^{+\infty}R(t-t')|A(t')|dt'\equiv\int_{-\infty}^{+\infty}R(t')|A(t-t')|dt'$ to expand the envelope in a Taylor series: $|A(t-t')|^{2}\simeq |A(t)|^{2}-t'\de_{t}|A(t)|^{2}+\frac{1}{2}t''\de_{t}^{2}|A(t)|^{2}+...$. By inserting this expression into Eq. (\ref{eq1}), we obtain the well-known Raman-NLSE model \cite{agrawalbook}: $i\de_{z}A+\frac{1}{2}s\de_{t}^{2}A+|A|^{2}A-TA\de_{t}|A|^{2}=0$, where the last term models the dynamics of long pulses subject to the Raman effect. In silica fibers this approximation is quite good for picosecond pulses, since $T\sim 2$ fsec, \cite{agrawalbook}. A completely different scenario occurs in the opposite limit, i.e. when $T\gg t_{0}$. In CS$_{2}$ liquid, $T\sim 1.68$ psec \cite{cs2}. In that case, one expands in a Taylor series the {\em response function} (which is varying slowly with respect to $A$):
$R(t-t')\simeq R(t)-t'\de_{t}R(t)+\frac{1}{2}t''\de_{t}^{2}R(t)+...$. Maintaining only the zero-th order term, Eq. (\ref{eq1}) becomes:
\eq{eq3}{i\de_{z}A+\frac{1}{2}s\de_{t}^{2}A+\mathcal{E}R(t)A=0,} where $\mathcal{E}\equiv\int_{-\infty}^{+\infty}|A(t)|^{2}dt$ is the total number of photons launched into the fiber. In the highly non-instantaneous limit,  the GNLSE (\ref{eq1}) is equivalent to a {\em linear} \schr equation with a time-dependent potential $R(t)$. In this regime, the response time is so long with respect to the pulse duration, that the system 'remembers' the {\em total} energy injected into the fiber, while the pulse is propagating. Consequently, the pulse feels the exponentially decaying response of the liquid as it were a purely linear potential,
whose depth is determined by the energy $\mathcal{E}$.

{\it Linearons states ---} Localized solutions of Eq. (\ref{eq3}) can be found by solving the equation for $t<0$ and $t>0$, and then by imposing continuity of the envelope and its derivative at $t=0$. These continuity conditions provide the constraint equation for the wavenumber $\beta>0$. The solution of (\ref{eq3}) with the exponential potential is known from standard quantum mechanics, being the exponential well among the solvable potentials.
Soliton solutions of Eq. (\ref{eq3}) are found by posing $A(z,t)=a(t)\exp\left(i\beta z\right)$, which leads to (for $t>0$)
\eq{eq5}{\frac{1}{2}\de_{t}^{2}a+\frac{\mathcal{E}}{T}\exp(-t/T) a=\beta a.}
For ($t<0$) the exponentially localized solution is
$a=\mathcal{N}\exp(\sqrt{2\beta} t)$ where $\mathcal{N}$ is a normalization constant, determined by the {\em total} energy $\mathcal{E}$.
Among the possible solutions of (\ref{eq5}) for $t>0$, we choose those vanishing as $t\rightarrow \infty$, which leads to \eq{eq7}{a=\mathcal{N}\frac{ J_\nu(\sqrt{8\mathcal{E}T} e^{-t/2T})}{J_\nu(\sqrt{8\mathcal{E}T})}}
with $\nu=\sqrt{8\beta}T$.
For large $t$, $a$ decays as $\exp(-\nu t/2 T)$.
The allowed eigenvalues of $\beta$ for a specified value of $\mathcal{E}$ are found by solving the implicit equation
 \eq{impl2}{_{0}F_{1}\left(;\sqrt{8\beta}T;-2\mathcal{E}T\right)=0,} where $_{0}F_{1}(;a;z)=\sum_{n=0}^{\infty}z^{n}/[n!(a)_{n}]$ is the confluent hypergeometric limit function, and $(a)_{n}\equiv \Gamma(a+n)/\Gamma(a)$ is the rising factorial.
Eq. (\ref{impl2}), which provides the dispersion relation $\beta(\mathcal{E},T)$, can only be solved numerically. The normalization constant
$\mathcal{N}$, which cannot be written explicitly, is found by requiring that the total soliton energy to equal $\mathcal{E}$. Solutions can be classified by counting the number of nodes ($m$). We shall call the 'fundamental' state the solution with $m=0$, which has the largest $\beta$ and the shortest pulse duration.
For small values of $\mathcal{E}$, Eq. (\ref{impl2}) can be expanded to give $\beta\simeq \mathcal{E}^{2}/2$, which shows that $\beta=\mathcal{E}=0$ is also a solution. For large values of $\mathcal{E}$, we have found that, for the fundamental state of largest $\beta$, one has the quasi-linear behavior $\beta\simeq (2/\pi)^{2}\mathcal{E}/T$.
Analytical solutions for moving solitons can also be found by taking the solutions of Eq. (\ref{eq5}) and making the substitution $2\beta\rightarrow 2\beta-v^{2}$, and by multiplying the fields by a factor $e^{-ivt}$, where $v$ is an extra parameter associated with the soliton velocity. In this case, we have an extra constraint $\beta>v^{2}/2$ that must be satisfied for a real solution to exist.
For $|t|$ large, the field decays as $a\sim\exp(-\sqrt{2\beta}|t|)$. This means that the original approximation used to derive Eq. (\ref{eq3}) from Eq. (\ref{eq1}) is verified {\em a posteriori} only if $\beta\gg [2T^{2}]^{-1}$, or by using the large energy limit $\mathcal{E}\gg\pi^{2}/(8T)$, which is the physical condition for our simplified highly-nonlocal model to be valid.

\begin{figure}
\includegraphics[width=8cm]{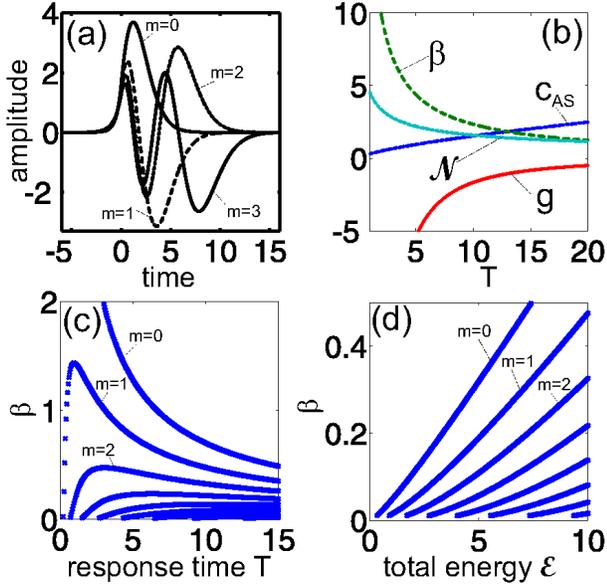}
\caption{(Color online) (a) Plots of localized solutions of Eq. (\ref{eq3}) for $T=10$, $\mathcal{E}=30$, for $\beta=2.33614$ ($m=0$), $1.82551$ ($m=1$), $1.45665$ ($m=2$) and $1.16691$ ($m=3$), calculated by using Eq. (\ref{impl2}). (b) $\mathcal{N}$, $c_{\rm AS}$, $g$, $\beta$ as functions of $T$, for $\mathcal{E}=30$. (c) Plot of $\beta(T)$ for fixed $\mathcal{E}=10$. (d) Plot of $\beta(\mathcal{E})$ for fixed $T=10$.
\label{fig1}}
\end{figure}

Figure \ref{fig1}(a) shows the profiles of some analytical solutions of Eq. (\ref{eq3}). Figures \ref{fig1}(b,c) show the dependence of $\beta$ on $\mathcal{E}$ and  $T$ for fixed $T$ and $\mathcal{E}$ respectively, as calculated by solving Eq. (\ref{impl2}) numerically. Approximate expressions in terms of $\mathcal{E}$ and $T$ are given below.

\begin{figure}
\includegraphics[width=8cm]{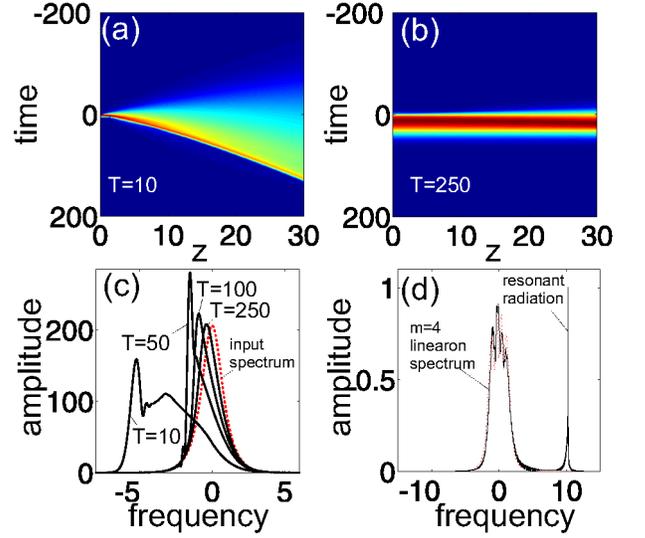}
\caption{(Color online)(a) Propagation of 1-soliton in full eq., $T=10$, $\mathcal{E}=30$, $\beta=2.33614$. (b) Same as (a) but for $T=250$, $\mathcal{E}=30$, $\beta=0.109846$, showing reduction of RSFS-induced acceleration. (c) Various final spectra showing reduction of RSFS ($\mathcal{E}=32$, $T=10,50,100,250$). (d) Resonant radiation in reduced model for $\alpha=0.05$ (for $m=4$ linearon). \label{fig2}}
\end{figure}

\begin{figure}
\includegraphics[width=8cm]{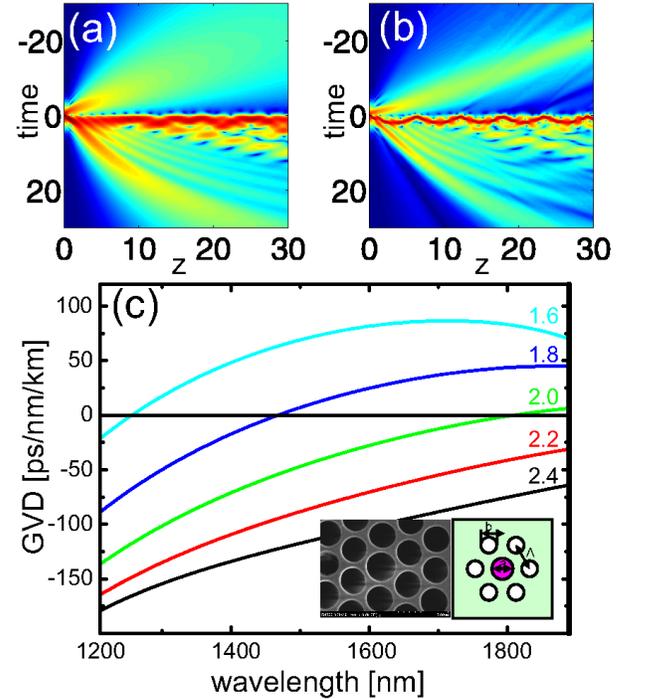}
\caption{(Color online) (a,b) Temporal oscillations of linearons due to a small instantaneous Kerr nonlinearity [$r=0$ in (a) and $r=1/6$ in (b), $T=10$, input pulse $A=N\sech{t}$ with $N=5$], in the reduced model of Eq. (\ref{eq3}).
(c)  GVD of the fundamental core mode (HE$_{11}$-mode) of a CS$_{2}$-filled PCF.
The structure consists of the liquid CS$_{2}$ core and one additional ring of air holes (see the inset), with $a = 1.6$ $\mu$m, $b = 1.2$ $\mu$m. Each curve refers to a structure of a different hole spacing $\Lambda$ expressed in $\mu$m. \label{fig3}}
\end{figure}

{\it Raman self-frequency shift of linearons ---} The analytical solutions of the reduced model (\ref{eq3}) calculated above, being linear, are stable against small noise fluctuations in their profile. To prove this, in Fig. \ref{fig2}(a) we show as an example the propagation of the higher-order soliton shown in Fig. \ref{fig1}(a) [blue line], when perturbing it with 5\% noise. If the same analytical solution is propagated inside the full model (\ref{eq1}), we observe that the pulse is subject to a Raman shift, which is not described by the simplified model of Eq. (\ref{eq3}). We have developed a general theory of RSFS of solitons for Eq. (\ref{eq1}). We rewrite Eq. (\ref{eq1}) as $i\de_{z}A+\frac{1}{2}s\de_{t}^{2}A+\mathcal{E}R(t)A+\left[\int R(t-t')|A(t')|^{2}dt'-\mathcal{E}R(t)\right]A=0$, and treat the term under square brackets as a perturbation of the stationary states found in the previous section.
One can prove that the central frequency of the soliton shifts according to $\Omega(z)=-(z/\mathcal{E})\int dt |A(t)|^{2}\de_{t}\left[\int R(t-t')|A(t')|^{2}dt'-\mathcal{E}R(t)\right]\simeq c_{\rm AS}z\int dt \left[\de_{t}^{2}|A(t)|^{2}\right]R(t)$, where $c_{\rm AS}\equiv\int t'|A(t')|^{2}dt'/\int |A(t')|^{2}dt'$ is the asymmetry coefficient of the localized state. The final result is ($x\equiv\sqrt{8\mathcal{E}T}$):
\begin{equation}
\begin{array}{l}
\Omega=\frac{c_{\rm AS}z \mathcal{N}^2}{J_{\nu}(x)^2 T}\int_{0}^{\infty}e^{-t/T}\de_{t}^{2}J_{\nu}^{2}(x e^{-t/2T})dt\cong -\frac{32}{\pi^7}\frac{\mathcal{E}^2}{T}z\text{,}
\label{RSFS}
\end{array}
\end{equation}
which shows that the RSFS of these solitons depends on the asymmetry of their temporal profiles. The last expression in (\ref{RSFS}) is valid for large $\mathcal{E}$ and $T$ (where it is possible to show  $\mathcal{N}^2\cong 2\mathcal{E}/\pi^2 T$ and $c_{AS}\cong T/\pi^2$),
and illustrates how, as the degree of nonlocality $T$ grows, the Raman shift is inhibited, while also growing with the energy.
The rate of RSFS $g\equiv\Omega/z$ and the asymmetry coefficient $c_{\rm AS}$ of the stationary states are shown in Fig. \ref{fig2} in terms of $T$ and $\mathcal{E}$ for the fundamental soliton solutions.

{\it Dispersive resonant radiation ---} In the presence of higher-order dispersion terms in Eq. (\ref{eq3}), one can show that linearons can resonantly emit dispersive radiation at well-defined frequencies, analogously to what occurs for \schr solitons in solid-core PCFs \cite{akhmediev}. By substituting $A(z,t)=[F(t)+f(z,t)]e^{i\beta z}$ into Eq. (\ref{eq3}), where $F(t)$ is the linearon profile and $f$ is the small dispersive radiation amplitude, assuming that the response time $T$ is large in comparison with the linearon duration, and keeping only the first order term in $f$, we have: $\left[i\de_{z}-\beta+\hat{D}(i\de_{t})+\mathcal{E}R(t)\right]f=-\hat{D}_{\rm H}(i\de_{t})F$, where $\hat{D}(i\de_{t})\equiv (1/2)\de_{t}^{2}+i\alpha\de_{t}^{3}$ and $\hat{D}_{\rm H}\equiv i\alpha\de_{t}^{3}$, and $\alpha$ is the third-order dispersion coefficient, the only one that we include here. At phase-matching, radiation and soliton have the same wavenumber $\beta$ and $f(\omega)\simeq S(\omega)/\left[D(\omega)+\mathcal{E}R(\omega)-\beta\right]$, where $S(\omega)$ is the Fourier transform of the source term $-\hat{D}_{\rm H}(i\de_{t})F$. This yields the phase-matching condition $D(\omega)+\mathcal{E}R(\omega)\simeq \beta$, which determines the resonant frequency $\omega_{\rm R}$. The energy-dependent part, although quite small for $T\rightarrow\infty$, is an extra contribution to the resonant condition that is unique for linearons, and allows the frequency position of the emitted radiation to be tuned by adjusting the {\em total} input pulse energy injected into the fiber.

{\it Kerr nonlinearity ---} An important issue concerns the effect of a residual instantaneous Kerr nonlinearity on the linear dynamics. This may be due
to the Kerr effect of the fiber cladding, or of the liquid itself. As shown in panels \ref{fig3}(a,b) for a
sech-like input pulse (i.e. an input that is not matched with the profile of the fundamental linearon), in the absence
of the Kerr effect [Fig. \ref{fig3}(a)], the propagation largely resemble a standard fundamental soliton. Conversely, increasing
the contribution of the Kerr effect, parameterized by a coefficient $r$ which is the relative importance between the instantaneous and the non-instantaneous part of the nonlinearity (typically of the order of $\sim 0.1$ in liquids) [Fig. \ref{fig3}(b)], induces a characteristic oscilation of the central position of the pulse in the temporal
and in the spectral domain. Such an effect is due to the coupling between higher order linearons that get
excited in the fiber, induced by the perturbation introduced by the Kerr nonlinearity. The details of such a perturbation theory will be reported elsewhere.

{\it Linearon-fiber ---}
In Fig. \ref{fig3}(c) we show the calculated GVD of a series of CS$_{2}$-filled core silica PCFs (with a single ring of holes arranged in a triangular cladding lattice as shown in the inset) for different values of the pitch $\Lambda$. Other parameters are given in the caption. The PCF introduces tremendous flexibility in the engineering of the position of the zero-GVD point, which can be shifted over a large range of wavelengths. This will allow optimal experimental excitation of linearons in the fiber, and the formation of linearon-induced supercontinua.

{\it Conclusions ---} We predict the existence of a novel class of temporally localized waves
propagating inside microstructured fibers with a central core filled by nonlinear liquids with a slow reorientational nonlinearity.
Surprisingly, these waves (which we dubb {\em linearons}) behave very much like solitons -- albeit being the solution of a linearized equation -- and are sustained by a strongly nonlocal temporal response, due to a pronounced Raman-like effect induced by reorientational nonlinearity of cigar-shaped molecules.
By borrowing concepts from nonlocal spatial solitons, we find that these objects are very robust with respect to
noise, caused by for example amplified spontaneous emission. Linearons may thus support the development of
novel classes of lightwave fiber systems and novel soliton based lasers, as well as opening up a new route
towards quantum solitons and multidimensional solitary waves sustained by a highly non-instantaneous nonlinearity.
In addition, by exploiting the nonlinear coupling between linearons, we believe that it will be possible to control and enhance supercontinuum generation in novel and unexpected ways.

C.C. acknowledges support from ERC Grant 201766. F.B., P.St.J.R. and M.S. are funded by the German Max Planck Society for the Advancement of Science (MPG).

\end{document}